\documentclass{cernrep}
\usepackage{texnames}
\usepackage[T1]{fontenc}

\usepackage{endnotes}       
\usepackage{color}
\newcommand{\ced}[1]{\textcolor{black}{#1}}

\def\u{\:}           
\usepackage{amsfonts}
 \DeclareSymbolFont{upgreek}{U}{eur}{m}{n}
 \DeclareMathSymbol{\rmalpha}{0}{upgreek}{"0B}
 \DeclareMathSymbol{\rmbeta}{0}{upgreek}{"0C}
 \DeclareMathSymbol{\rmgamma}{0}{upgreek}{"0D}
 \DeclareMathSymbol{\rmdelta}{0}{upgreek}{"0E}
 \DeclareMathSymbol{\rmmu}{0}{upgreek}{"16}
 \DeclareMathSymbol{\rmpi}{0}{upgreek}{"19}
 \DeclareMathSymbol{\rmsigma}{0}{upgreek}{"1B}
 \DeclareMathSymbol{\rmphi}{0}{upgreek}{"1E}
 \DeclareMathSymbol{\rmomega}{0}{upgreek}{"21}

\begin{document}
\title{Microwave Discharge Ion Sources}

\author{L. Celona}

\institute{Istituto Nazionale di Fisica Nucleare, Laboratori Nazionali del Sud, Catania, Italy}

\maketitle 

\begin{abstract}
This \ced{chapter} describes the basic principles, design features and characteristics of microwave discharge ion sources. A suitable source for the production of intense beams for high-power accelerators must \ced{satisfy the requirements of} high brightness, stability and reliability. The 2.45{\u}GHz off-resonance microwave discharge sources are ideal devices to generate the \ced{required} beams, as they produce multimilliampere beams of protons, deuterons and \ced{singly charged} ions. A description of different technical designs will be given, analysing their performance, with particular \ced{attention being paid} to the  quality of the beam, especially in terms of its emittance.
\end{abstract}

\section{Introduction}

The production of high-current beams is a key \ced{requirement for various} applications, and this \ced{is expected} to increase in coming years, either for industrial applications or for research projects. High-current and high-brightness
H$^+$ beams can be provided by microwave discharge ion
sources \ced{(MDISs)}, which present many advantages in terms of compactness,
high reliability, ability to operate in continuous-wave (CW) or pulsed
mode, reproducibility and low maintenance. Some applications based on intense proton beams  \cite{handbook} are \ced{as follows}:
\begin{itemize}
\item accelerator-driven systems (ADSs) for nuclear waste transmutation  and energy production,
\item radioactive ion beams,
\item intense neutron spallation sources,
\item radiation processing, and
\item neutrino factory.
\end{itemize}

The main parameters of the related proton drivers are listed in Table \ref{ProtonDriver}, while Table \ref{accelerator} shows a list of projects (operating and under construction) using high-current proton beams or intense H$^-$ sources with low transverse emittance.

\begin{table}[!h]
\caption{Proton driver requirements.}
\begin{center}\label{ProtonDriver}
\begin{tabular}{lcc}
\hline\hline
\textbf{Proton driver} & \textbf{Energy (GeV)} & \textbf{Beam power (MW)} \\
\hline
ADS: XADS & $\sim{0.6}$& $\sim{5} $\\
 \phantom{ADS:}       Ind. burner &  $\sim{1}$                 &$\sim{50}$\\
Spall. neutron source (ESS)& 1.33 & 5\\
Irradiation facility&$~ 1$&$>10$\\
Neutrino factory (CERN)&$2.2$&$4$\\
RIB: `one stage' &$\sim{0.2} $&$\sim{0.1}$\\
\phantom{RIB:} `two stage' &$\sim{1}$& $\sim{5}\mbox{--}10$\\
\hline\hline
\end{tabular}
\end{center}
\end{table}

\begin{table}[!h]
\caption{High-power accelerator requirements.}
\begin{center}\label{accelerator}
\begin{tabular}{lcccccc}
\hline\hline
& \textbf{p/H$^-$} & \textbf{mA} & \textbf{ms} & \textbf{Hz} & \textbf{Duty factor (\%)} & \textbf{\boldmath$\rmpi${\u}mm{\u}mrad}\\
\hline
LEDA    &  p     &  100  &  CW  &  CW  &  100  &  0.25\\
IPHI    &  p     &  100  &  CW  &  CW  &  100  &  0.25\\
TRASCO  &  p     &  30   &  CW  &  CW  &  100  &  0.2\\
ESS     &  p     &  60/90  &  2.84  &  14  &  4  &  0.3\\[6pt]
SPL     &  H$^-$ &  50  &  1.5  &  50  &  7.5  &  0.2\\
SNS     &  H$^-$ &  50  &  1  &  60  &  6  &  0.25\\
JKJ     &  H$^-$ &  30  &  0.5  &  50  &  2.5  &  0.25\\
\hline\hline
\end{tabular}
\end{center}
\end{table}

The optimization of beam formation and transport through the low-energy beam transport (LEBT) plays a fundamental role in \ced{the provision of a high-quality beam to the accelerator}. This is a common \ced{requirement} of the projects reported in Table \ref{accelerator}, where emittances at the entrance to the radio frequency quadrupole (RFQ) of the order of 0.20--0.30{\u}$\rmpi${\u}mm{\u}mrad are needed, making \ced{it essential to design and test} the ion source and LEBT as a
whole. The major challenge of the accelerator
front-end is therefore the preparation of a high-quality beam, \ced{with a pulse
that is well defined in time and has a small transverse emittance}.
In the following, a review of the major experiences in the production of intense proton beams are reported (Table \ref{accelerator}) together with future perspectives.

\section{Microwave ion source for high-intensity proton production}

\subsection{Historical notes}

\begin{figure}[htpb]
\centering
\includegraphics[width=0.5\textwidth]{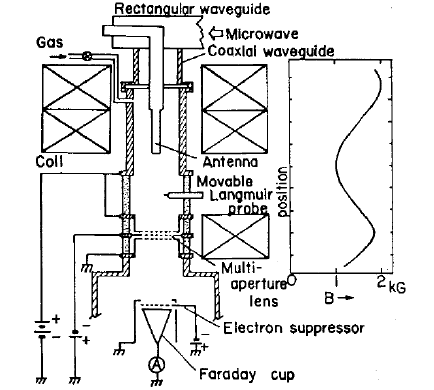}
\hspace{3mm}
\includegraphics[width=0.5\textwidth]{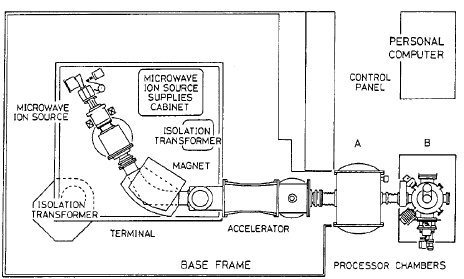}
\caption{Hitachi microwave ion source with multiaperture extraction
system\label{SakudoSource}}
\end{figure}

The history of 2.45{\u}GHz high-current source (HCSs) started about 35 years ago
with different source designs proposed by Sakudo \cite{Sakudo} and by
Ishikawa \emph{et al.} \cite{Ishikawa}, especially for industrial applications. The
sources produced remarkable results not only for protons, but
also for deuterons and \ced{singly charged} light ions. A simple concept
of the microwave discharge source was based on a non-confining
magnetic field higher than the resonance field (i.e.\ 87.5{\u}mT).
Sakudo and his collaborators at the
Central Research Laboratory of Hitachi Limited pioneered
the development of high-current microwave ion sources for
ion implantation \cite{brown}. The first Hitachi ion source (see Fig.~\ref{SakudoSource}) is composed of a  plasma generator, that is, essentially, a
section of coaxial waveguide with an axial magnetic field
supplied by three solenoids. The 2.45{\u}GHz microwaves are
introduced via a water-cooled antenna connected to the
inner conductor of a coaxial-to-rectangular waveguide
transition. The magnetic induction is varied along the
length of the antenna to match the impedance of the plasma-filled chamber to the impedance of the microwave
line \cite{Taylor}. The extraction system is a multiaperture triode with
124 apertures 3{\u}mm in diameter distributed over a 50{\u}mm
diameter circle.
The sources built by Sakudo's group  were able to supply 2{\u}mA of As$^+$ and 15{\u}mA of B$^+$, and they were successfully
adapted to industrial application setups.

\begin{figure}[htpb]
\begin{center}
\includegraphics[width=0.5\textwidth]{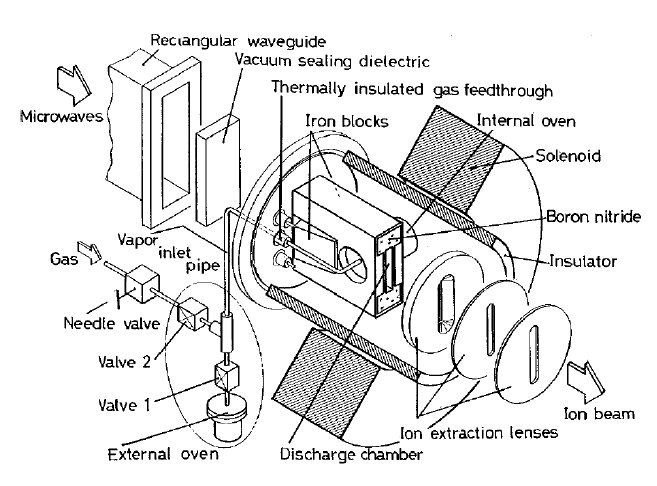}
\end{center}
\caption{Hitachi microwave ion source with slit extraction system
\label{IonImplanter}}
\end{figure}

The second microwave ion source
developed by Sakudo \emph{et al.} was especially designed to
generate a slit-shaped ion beam. The plasma chamber illustrated
in Fig.~\ref{IonImplanter} is a tapered ridged waveguide with all
but the volume between the ridges filled with boron nitride.
The 2.45{\u}GHz microwaves are introduced through a dielectric
window from a rectangular waveguide. The electric
field between the ridges is relatively uniform, ensuring a
reasonably constant plasma density over the entire length.

A step forward was made by Ishikawa \cite{Ishi}, whose
design was very compact (chamber diameter was 50{\u}mm) and the source was able to produce milliampere beams of any
species, finding applications not only in ion implantation devices
but also for ion beam deposition. The absence of antennas
made this equipment more reliable for long-time operations.

\subsection{The CRNL ion source}

\begin{figure}[htpb]
\begin{center}
\includegraphics[width=0.4\textwidth]{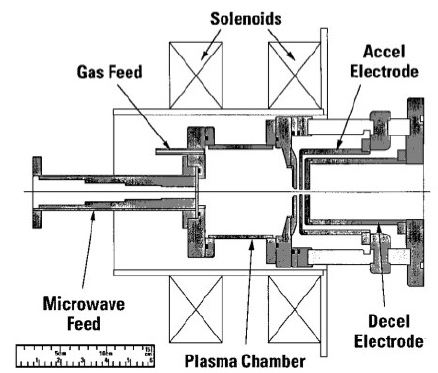}
\end{center}
\caption{The CRNL microwave source\label{ChalkRiver}}
\end{figure}

In 1991 a simple and robust design was proposed
by Taylor and Mouris at Chalk River National Laboratory (CRNL; see Fig.~\ref{ChalkRiver}). This source can be considered as the basis of all
the different designs proposed in the past 25 years.
The main innovation consisted in the use of a matching
unit to adapt the waveguide to plasma impedance, which
enhanced the plasma density and finally the current density
of the extracted beam. Moreover, two separately fed solenoids,
approximately placed at the two extremes of the
plasma chamber, \ced{allowed the magnetic field profile to be adapted}.
The extraction system was not sophisticated but adequate
for high-current beam formation, using the three electrodes
in the \ced{accelerate--decelerate}
operation mode.

\begin{figure}[htpb]
\begin{center}
\includegraphics[width=0.6\textwidth]{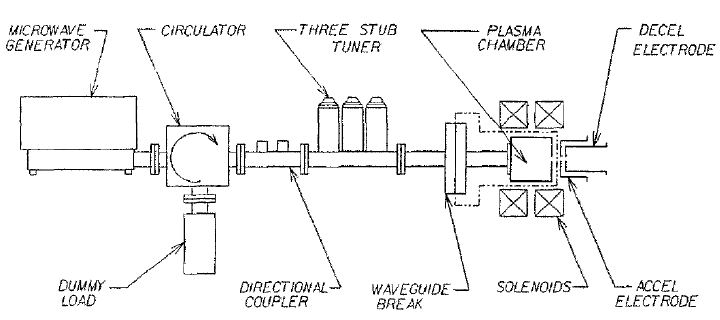}
\end{center}
\caption{The CRNL microwave source\label{ChalkRiverECRprotonsource}}
\end{figure}

The configuration of the Chalk River \ced{electron cyclotron resonance} (ECR) proton
source is illustrated in Fig.~\ref{ChalkRiverECRprotonsource}. The plasma generator is
simply a hydrogen-filled chamber, with a ceramic rectangular
waveguide window, encircled by two solenoids. The
extraction system is a 50{\u}kV multiaperture triode.

\begin{figure}[htpb]
\begin{center}
\includegraphics[width=1\textwidth]{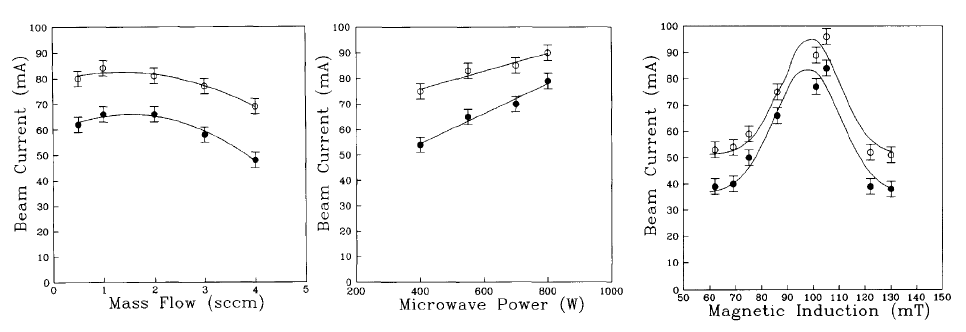}
\end{center}
\caption{Beam current as a function of mass flow, microwave power and magnetic induction \label{TaylorWills}}
\end{figure}

The microwave line is isolated from the plasma chamber by a d.c.\
waveguide break consisting of a Teflon sheet sandwiched
between a choke flange and a standard flange. Since the
solenoids are also electrically isolated from the plasma
chamber, only the plasma chamber itself remains at high
voltage. All of the power supplies, d.c.\ as well as microwave,
are at ground potential.
The ion source generates beam current densities of up
to 350{\u}mA{\u}cm$^{-2}$ at a microwave power of 1000{\u}W. The three-stub tuner is adjusted to
optimize the impedance match. Figure \ref{TaylorWills} illustrates the evolution of the beam current with the mass flow, microwave power and magnetic induction on the plasma chamber axis.

\subsection{The LEDA injector}

\begin{figure}[htpb]
\begin{center}
\includegraphics[width=0.6\textwidth]{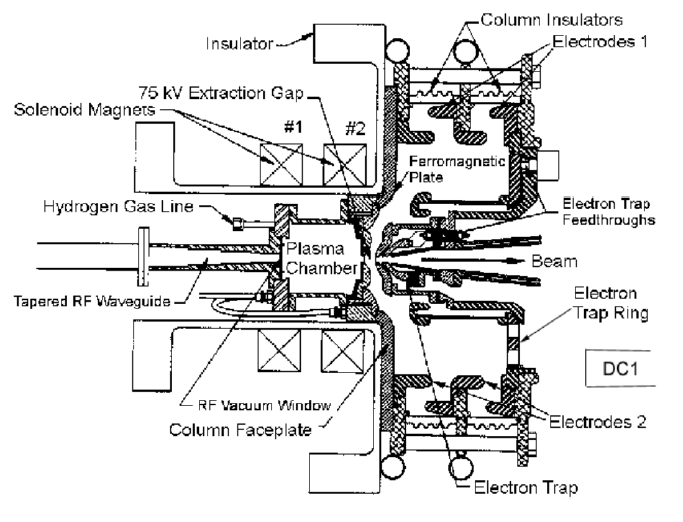}
\end{center}
\caption{Line drawing of the 75{\u}keV microwave proton source\label{LosAlamosAccelerator}}
\end{figure}

This design was further improved by J. Sherman and co-workers at Los Alamos National Laboratory (LANL), who improved the extraction system and the LEBT in order
to optimize the beam coupling to the RFQ of the LEDA project.
Then the CRNL plasma generator was integrated with a 75{\u}kV acceleration structure at Los Alamos (Fig.~\ref{LosAlamosAccelerator}) \cite{Sherman}. Plasma is generated by the interaction of 2.45{\u}GHz
microwaves with H$_2$ gas in the presence of an approximate
875{\u}G axial magnetic field.

\begin{figure}[htpb]
\begin{center}
\includegraphics[width=0.5\textwidth]{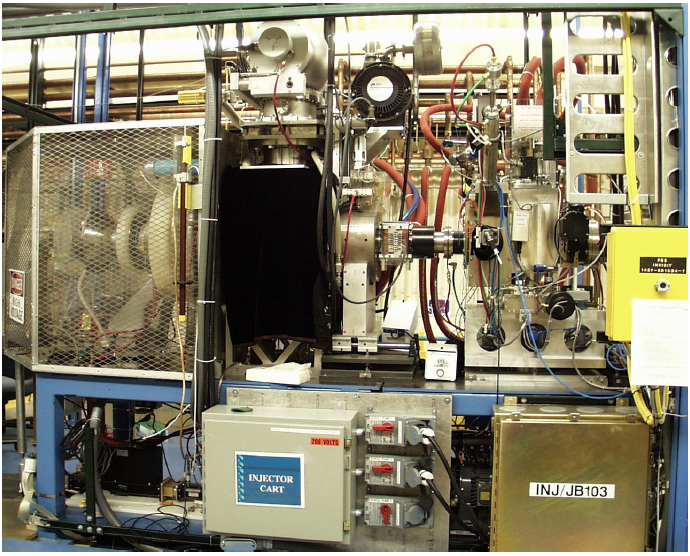}
\end{center}
\caption{Injector used for the beam measurements\label{inj_real}}
\end{figure}

Beam measurements
were made on a prototype injector shown in Fig.~\ref{inj_real}. The
proton source is mounted on a beam diagnostics and
pump box, which is followed by two solenoids. The beam
current is characterized by two d.c.\ current monitors (DC1
and DC2), an a.c.\ beam current transformer for beam noise measurements,
an $x$ and $y$ video profile system for beam position
and width measurements, and an emittance measuring unit (EMU), which also serves as the beam dump. The typical operating parameters of the injector are listed in Table \ref{inj_par}.

\begin{table}[!h]
\caption{LEDA injector parameters.}
\begin{center}\label{inj_par}
\begin{tabular}{lc}
\hline\hline
\textbf{Injector parameter}          & \textbf{Value}\\
\hline
H$_2$ gas flow                       & 4.1{\u}sccm*\\
Ion source pressure                  & 2{\u}mTorr\\
Ion source gas efficiency            & 24\% \\
Discharge power, at 2.45{\u}GHz      & 1.2{\u}\ced{kW}\\
Beam energy                          & 75{\u}keV\\
High-voltage power supply current    & 165{\u}mA\\
DC1 current                          & 154{\u}mA\\
DC2 current                          & 120{\u}mA\\
Proton fraction                      & 90\%\\
Injector emittance, 1 r.m.s.\ norm.  & 0.18{\u}$\rmpi${\u}mm{\u}mrad\\
\hline\hline
\multicolumn{2}{l}{\ced{\footnotesize{*sccm = standard cubic centimetres per minute.}}}
\end{tabular}
\end{center}
\end{table}

\begin{figure}[htpb]
\begin{center}
\includegraphics[width=0.6\textwidth]{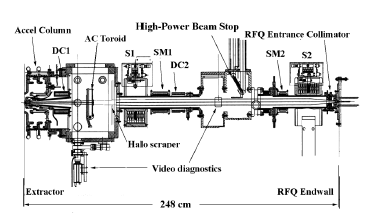}
\end{center}
\caption{Injector used for the beam measurements\label{top_view_inj}}
\end{figure}

From Fig.~\ref{top_view_inj} we can see that beam current is measured at DC1,
DC2 and in the RFQ entrance collimator for 6.7{\u}MeV RFQ; beam focusing accomplished
with LEBT is achieved thanks to solenoid magnets S1 and S2; and beam centroid  is controlled with steering magnets SM1 and SM2.

\subsection{The SILHI source}

\begin{figure}[htpb]
\begin{center}
\includegraphics[width=0.4\textwidth]{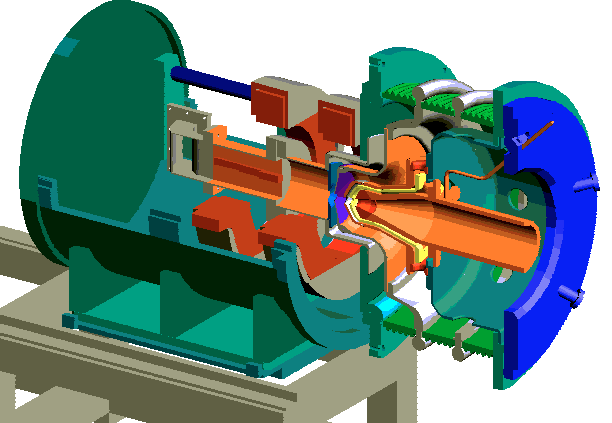}
\end{center}
\caption{Render view of the SILHI source operating at CEA\label{Silhi}}
\end{figure}

\begin{figure}[htpb]
\begin{center}
\includegraphics[width=0.8\textwidth]{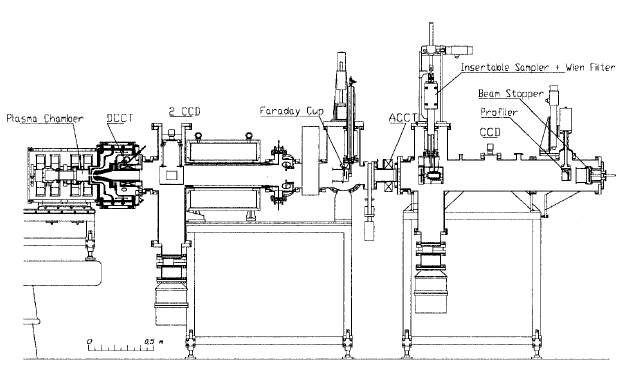}
\end{center}
\caption{Layout of the SILHI ion source and the beam line\label{silhi_layout}}
\end{figure}

Following the same track, at CEA-Saclay (Commissariat \`a
l'\'Energie Atomique, Saclay), the SILHI source obtained large brightness and high reliability in the second half of the 1990s. \ced{All the parts of the source, which is shown in Figs.~\ref{Silhi} and \ref{silhi_layout}, are robustly engineered, including the} beam line elements and diagnostics, adapted to beams
exceeding 10{\u}kW, typically 95{\u}kV,~140{\u}mA. A proton/deuteron fraction above 80\% and an emittance below
0.2{\u}$\rmpi${\u}mm{\u}mrad have been obtained, while the reliability \ced{has increased
over the years}, even reaching 99.9\% in a one-week test.
The SILHI source (see Fig.~\ref{silhi_layout}) has been developed at CEA-Saclay, in the framework of the
IPHI project, devoted to the production and acceleration
of intense proton beams (currents up to 100{\u}mA). The source
design and results are described in detail elsewhere \cite{RSI, lagniel,Gobin}, and in Table \ref{par_SILHI} we report the main features.

\begin{table}[htpb]
\caption{SILHI source typical operating parameters.}
\begin{center}\label{par_SILHI}
\begin{tabular}{lc}
\hline\hline
\textbf{Parameter}                  & \ced{\textbf{Value}}\\
\hline
Total beam current                  & 140{\u}mA\\
Proton fraction                     & $\sim{85}\%$\\
Beam density                        & 220{\u}mA{\u}cm$^{-2}$\\
Beam energy                         & 95{\u}keV\\
Discharge RF power, at 2.45{\u}GHz  & 1.2{\u}kW\\
Beam emittance, at 75{\u}mA         & 0.1{\u}\ced{$\rmpi${\u}mm{\u}mrad}\\
Hydrogen mass flow                  & $\sim{2}${\u}sccm\\
\hline\hline
\end{tabular}
\end{center}
\end{table}

A collaboration \ced{arranged} between CEA
and the Istituto Nazionale di Fisica Nucleare (INFN) \ced{has allowed in recent} years a significant increase of beam quality produced by SILHI, reaching the outstanding value of 0.1{\u}$\rmpi${\u}mm{\u}mrad for 0.75{\u}mA as described in detail in the next section.

\subsection{TRASCO Intense Proton Source (TRIPS) at INFN-LNS}

\ced{The TRIPS ion source \cite{celona} has been designed, realized and commissioned at INFN-LNS, within the framework of the TRASCO Project (an R\&D programme, the goal of which was the design of an accelerator driving system (ADS) for nuclear waste transmutation).} With reference to the \ced{three experiments outlined above}, a series of innovations were implemented on the TRASCO project. The
goal of TRIPS was less demanding in
terms of currents than it was for the other projects in Table \ref{accelerator}, but there was a \ced{very strong requirement for source reliability}, with an r.m.s.\ normalized emittance below 0.2{\u}$\rmpi${\u}mm{\u}mrad for an operating voltage of 80{\u}kV (Table~\ref{trip_req}).

\begin{table}[!h]
\caption{TRIPS typical operating parameters.}
\begin{center}\label{trip_req}
\begin{tabular}{lc}
\hline\hline
\textbf{Parameter}                  & \textbf{\ced{Value}}\\
\hline
Total beam current                  & 60{\u}mA\\
Proton fraction                     & 90\% \\
Beam energy                         & 80{\u}keV\\
Microwave power, at 2.45{\u}GHz     & 0.3--1{\u}kW\\
Beam emittance, at 35{\u}mA         & 0.07{\u}$\rmpi${\u}mm{\u}mrad\\
Gas flow                            & 0.4--0.6{\u}sccm\\
Duty factor, at 35{\u}mA            & 99.8\% \\
Reliability, at 35{\u}mA            & 99.8\%\\
\hline\hline
\end{tabular}
\end{center}
\end{table}

The major innovation
consisted in the use of two movable coils permitting
the magnetic field profile to be varied. The plasma chamber dimension
and the four-step matching transformer were defined to
get a uniformly dense plasma at the extraction hole (6 or 8{\u}mm diameter).
The microwave coupling with the matching
transformer and the automatic tuning unit allowed operation
with low values of reflected power (below 5\%) and a
high electric field on the axis, thus increasing proton fraction
and current density, up to 200{\u}mA{\u}cm$^{-2}$. The extraction system
was designed in collaboration with Saclay, exploiting the
experience already gained at INFN-LNS for the design of extraction geometries from plasma sources. Measurements with
the 8{\u}mm hole plasma electrode have given results largely
exceeding the TRASCO design current: up to 61{\u}mA were
extracted at 80{\u}kV and about 90\% of the beam has been transported to the beam stop. 

\begin{figure}[htpb]
\begin{center}
\includegraphics[width=0.7\textwidth]{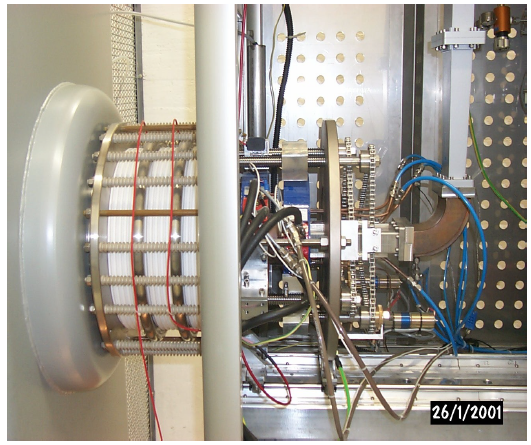}
\end{center}
\caption{The TRIPS ion source on the 100{\u}kV
platform\label{trip_source}}
\end{figure}

\begin{figure}[htpb]
\begin{center}
\includegraphics[width=0.7\textwidth]{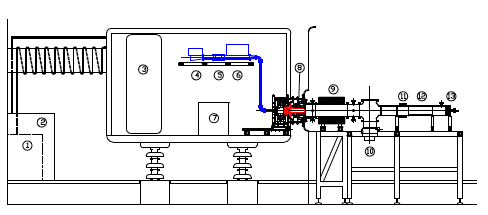}
\end{center}
\caption{The experimental set-up: (1) demineralizer; (2) 120{\u}kV insulating transformer; (3) 19{\u}inch rack for the power supplies and for the remote control system; (4) magnetron and circulator; (5) directional coupler; (6) automatic tuning unit; (7) gas box; (8) DCCT1; (9) solenoid; (10) turbomolecular pump; (11) DCCT2; (12) quartz tube; (13) 10{\u}kW beam stop. \label{trasco_setup}}
\end{figure}

Figure \ref{trip_source} shows the TRIPS source and Fig. \ref{trasco_setup} shows the experimental set-up.
The first section of the low-energy beam transfer line (LEBT) devoted to beam analysis consists of a
current transformer (DCCT1), a focusing solenoid, a four-sector ring to measure beam
misalignments and inhomogeneities, a second current transformer (DCCT2) and an insulated 10{\u}kW beam stop (BS), which measures the beam current.

\begin{figure}[htpb]
\begin{center}
\includegraphics[width=0.8\textwidth]{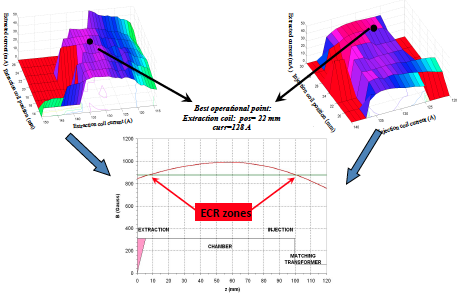}
\end{center}
\caption{Magnetic field optimization through changes in the coil current and position (above) and ideal profile (below).\label{optimum}}
\end{figure}

Systematic measurements with movable coils have been carried out for
different positions and currents, by leaving the other parameters
unchanged. From these measurements \ced{it can be determined} that the source is more
\ced{sensitive} to variations in the extraction coil than in the injection
one. The best performance is clearly obtained when
the two ECR zones are located exactly on the \ced{boron nitride} discs at the
two ends of the plasma chamber.
Figure \ref{optimum} shows the best obtained
profile (the straight line represents the value of the
magnetic field corresponding to the resonance at 2.45{\u}GHz).

Systematic emittance measurements have been made with an emittance measurement unit provided by the CEA-Saclay SILHI group, showing the effect of each parameter on the beam emittance; more details are reported in the next section.

\subsection{Versatile Ion Source (VIS)}

\begin{figure}[htpb]
\begin{center}
\includegraphics[width=0.4\textwidth]{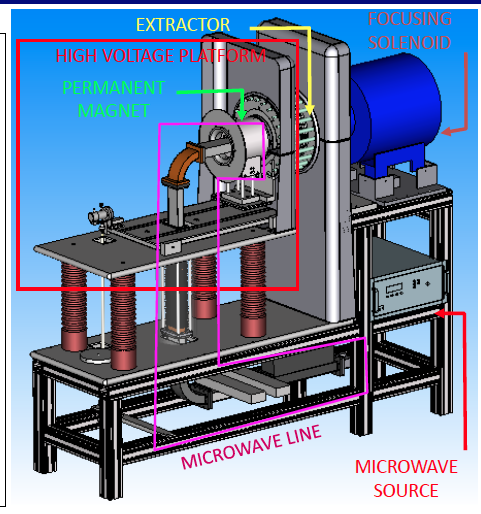}
\end{center}
\caption{A render view of the VIS source together with the focusing solenoid\label{VIS}}
\end{figure}

The TRIPS source \ced{exceeded} all the requirements of the TRASCO project,
but in order to simplify it, the VIS source was
built in 2006 \cite{cia}. In particular, in the VIS
source, the movable coils have been replaced with permanent magnets, and
the extraction geometry and extraction column have been simplified.
With these modifications it has been possible to avoid the high-voltage platform and the insulating transformer, by insulating the gas pipe and the waveguide line.
All these changes decreased the high-voltage sparks and
increased the source reliability. All the devices for the remote control were placed at
ground potential, thus leaving only the plasma chamber and the permanent
magnets at high voltage. The compact dimensions have also \ced{resulted in}
better and easier maintenance.
The magnetic system consists of a set of three
Vacodym 745~HR permanent magnets;
they are packed together with two
soft iron spacers and they are supported by a stainless-steel tube. The layout of the
VIS source is reported in Fig.~\ref{VIS}~\cite{vis}.

\begin{figure}[htpb]
\begin{center}
\includegraphics[width=0.3\textwidth]{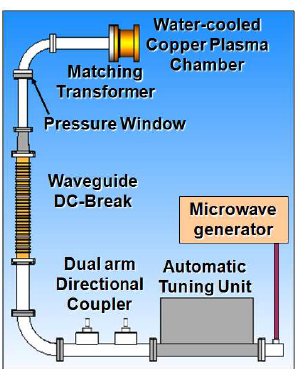}
\end{center}
\caption{VIS microwave line\label{micro_line}}
\end{figure}

The microwave line, in Fig.~\ref{micro_line}, is the result of an optimization study
carried out with tools for high-frequency structure simulation in order to
reduce microwave losses, simultaneously with an adequate matching
of the waves to the plasma chamber. A plasma is generated by means
of the microwaves provided by a 2.45{\u}GHz magnetron through a WR~340
(86.4{\u}mm $\times$ 43.2{\u}mm) waveguide excited in the TE$_{10}$ dominant mode. An
automatic tuning unit adjusts the modulus and phase of the incoming wave
in order to match the plasma chamber impedance with and without the
plasma, and a dual-arm directional coupler is used to measure the forward
and the reflected power.

\begin{figure}[htpb]
\begin{center}
\includegraphics[width=0.5\textwidth]{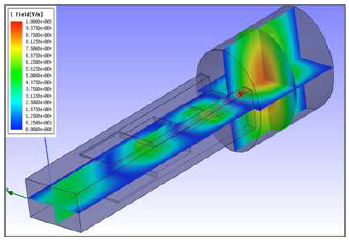}
\end{center}
\caption{Electric field amplitude at 2.45{\u}GHz\label{trasf}}
\end{figure}

In TRIPS both the microwave line and the microwave generator are
placed on the high-voltage platform, while in VIS the microwave generator
is placed at ground potential. Then, in order to separate the high-voltage
region from the grounded one, a waveguide d.c.\ break has been designed and
realized with the support of the HFSS electromagnetic simulator.
It is made of 31 aluminium discs of a WR~340 waveguide insulated \ced{from one another}
by means of fibreglass. The conductive parts will be fixed
to voltages gradually decreasing from 80{\u}kV to ground voltage, still keeping
the insertion loss low.
The high-pressure quartz window is placed before the WR~284 water-cooled
copper bend in order to avoid any damage due to back-streaming
plasma electrons. Finally, a maximally flat matching transformer has been
inserted before the plasma chamber, shown in Fig.~\ref{trasf}, which
is an optimized version of a similar device used in MIDAS and TRIPS ion
source.
This realizes an impedance matching with the plasma chamber and
concentrates the electromagnetic field near the axis. The enhancement of
the electromagnetic field in the plasma chamber cavity increases the
plasma density locally and finally the ionization process.

\begin{figure}[htpb]
\begin{center}
\includegraphics[width=0.3\textwidth]{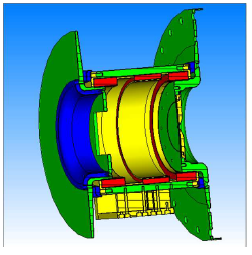}
\end{center}
\caption{Magnet\label{Magnet}}
\end{figure}

\begin{figure}[htpb]
\begin{center}
\includegraphics[width=0.4\textwidth]{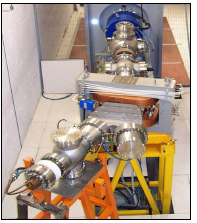}
\end{center}
\caption{VIS beam line\label{testbench}}
\end{figure}

The ionic component of the plasma produced in the plasma chamber is then
extracted by means of a four-electrode extraction system. The low-energy
beam transport line (LEBT) allows beam analysis, and it consists of a
focusing solenoid, a four-sector diaphragm to measure the beam misalignments,
a d.c.\ current transformer, a 30$^\circ$ bending magnet (Fig.~\ref{Magnet}) and an insulated beam stop to measure the beam current, shown in Fig.~\ref{testbench}.
The extraction system consists of a plasma electrode made of molybdenum
at 65{\u}kV voltage, two water-cooled grounded electrodes and a 3.5{\u}kV
negatively biased screening electrode inserted between them, in order to
avoid secondary electrons due to residual gas ionization, back-streaming to
the extraction area. The VIS extraction has been optimized to work around
40{\u}mA, fulfilling
the requirement of high brightness. The beam emittance measurements are described in the next section.
 The beam availability has been further increased and
damage to the electronics because of high-voltage sparks is negligible.

\section{Beam emittance measurements}

The production of a high-quality beam has been, since the first measurements made at CRNL by Taylor and Wills, one of the major challenges for such sources, together with beam reliability. Over the years, many developments have been carried out to improve these aspects, and the most important results have been obtained through a deep optimization of the extraction geometry together with an appropriate design of the low-energy  beam transfer line.

\subsection{SILHI beam emittance improvements}

\ced{A significant step forward in this field, namely the production of a high-quality beam, was made within the framework of the collaboration between INFN and CEA,} in which an innovative method based on the controlled injection of a gas into the line was developed \cite{gob}. The idea is based on the fact that the ions obtained from residual gas ionization are expelled from
the centre of the beam line, where the potential is positive, towards the wall.
Electrons from the wall are attracted \ced{towards}
the beam, so that the beam is compensated, provided that the pressure is high enough to have an adequate number of electrons (a compromise between beam losses and space-charge
compensation is to be found experimentally). According to that approach,
the most effective gases are the heaviest, which easily
release a large number of electrons.
\ced{If $N$ hydrogen atoms per unit volume are
required for optimum compensation} (i.e.\ if $N$ electrons
neutralize the beam space charge), for a species that gives
$Z_{\rm eff}$ electrons when the atoms interact with the 95{\u}keV proton
beam, the optimum number of atoms per unit volume is $N/Z_{\rm eff}$ (this clue neglects the dissociation process, which does not require a large amount of energy).

Experiments have been carried out by injecting different gases (H$_2$, N$_2$, $^{84}$Kr and Ar) into the SILHI beam line through a leak valve placed after the LEBT solenoid and comparing the emittance measurements at different pressures with an extracted beam in range of 75--80{\u}$\rmmu$A. Two gauges (\ced{measuring} $p_1$ and $p_2$) are placed, respectively, at the extraction column exit and between the sampler and the
profiler in the diagnostic box (Fig.~\ref{silhi_layout}).
In all the cases considered here, a decrease of beam
emittance has been observed with the beam line pressure
increase. These results have been explained by a \ced{higher degree of
space-charge compensation} as confirmed by a series of measurements carried out in collaboration with the Los Alamos National Laboratory (LANL).

\begin{figure}[htpb!]
\centering
\includegraphics[width=10cm]{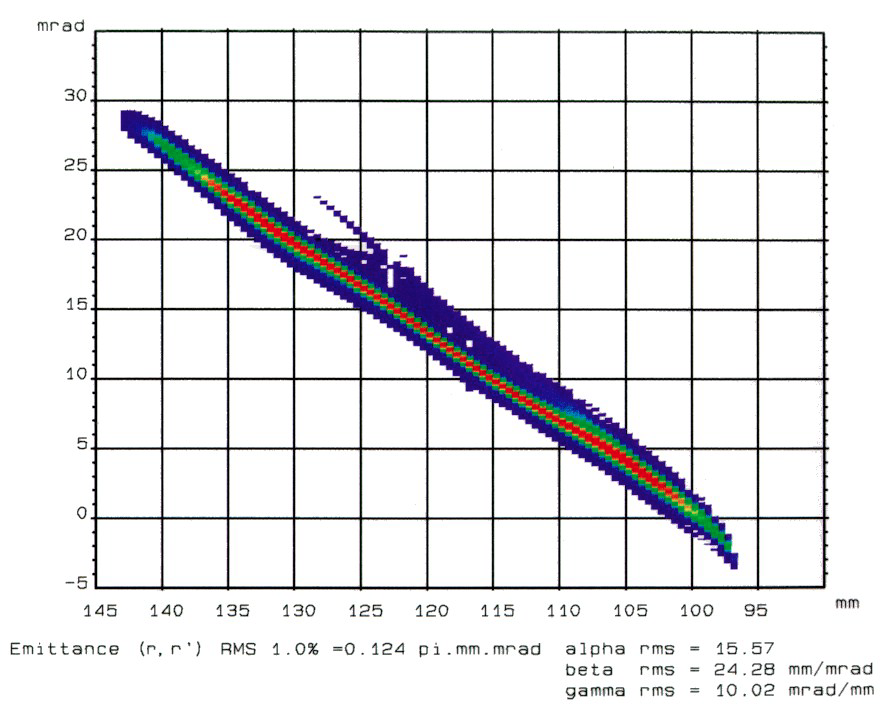}
\caption{Emittance plot after injecting Ar into the beam line:
$p_1=4.5\times10^{-5}${\u}Torr, $p_2=4.4\times10^{-5}${\u}Torr, $\epsilon_{\rm rms} = 0.124${\u}$\rmpi${\u}mm{\u}mrad. \label{w_Ar}}
\end{figure}

\begin{figure}[htpb!]
\centering
\includegraphics[width=10cm]{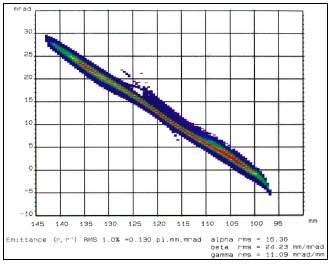}
\caption{Emittance plot after injecting N$_2$ into the beam line:
$p_1=4.5\times10^{-5}${\u}Torr, $p_2=4.5\times10^{-5}${\u}Torr, $\epsilon_{\rm rms} = 0.13${\u}$\rmpi${\u}mm{\u}mrad. \label{w_N2}}
\end{figure}

\begin{figure}[htpb!]
\centering
\includegraphics[width=9cm]{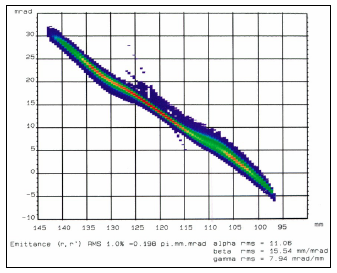}
\caption{Emittance plot after injecting H$_2$ into the beam line:
$p_1=5\times10^{-5}${\u}Torr, $p_2=4.9\times10^{-5}${\u}Torr, $\epsilon_{\rm rms} = 0.198${\u}$\rmpi${\u}mm{\u}mrad. \label{w_H2}}
\end{figure}

As observed, the behaviour depends upon the
atomic mass of the species injected. In particular, an r.m.s.\ normalized emittance of 0.1{\u}$\rmpi${\u}mm{\u}mrad has been obtained with $^{84}$Kr injected into the beam line ($p_1 = 3 \times 10^{-5}${\u}Torr). The emittance is reduced by a factor of~3 with
respect to \ced{the value 0.33{\u}$\rmpi${\u}mm{\u}mrad} measured without gas ($p_1 = 1.8 \times 10^{-5}${\u}Torr). Emittance values below
0.15{\u}$\rmpi${\u}mm{\u}mrad have been easily obtained also using Ar
with pressure values of the first gauge around $2.5 \times 10^{-5}${\u}Torr. For this gas, the minimum value of r.m.s.\ normalized
emittance of 0.125{\u}$\rmpi${\u}mm{\u}mrad has been measured (see Fig.~\ref{w_Ar}). A lower efficacy is obtained \ced{with} N$_2$ and H$_2$ injection: in particular, with N$_2$ injection we have measured 0.13{\u}$\rmpi${\u}mm{\u}mrad at relatively
high pressure ($p_1 = 4.5 \times 10^{-5}${\u}Torr; see Fig.~\ref{w_N2}), while for H$_2$ the minimum value of emittance obtained is 0.198{\u}$\rmpi${\u}mm{\u}mrad,
as shown in Fig.~\ref{w_H2}.

\begin{figure}[htpb!]
\centering
\includegraphics[width=10cm]{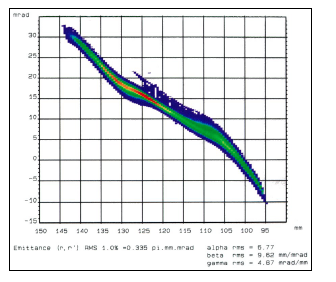}
\caption{Emittance picture without injecting $^{84}$Kr into the beam line: $\epsilon_{\rm {rms}} = 0.33${\u}$\rmpi${\u}mm{\u}mrad, $p_1=8\times10^{-5}${\u}Torr, $p_2=1.2\times10^{-5}${\u}Torr. \label{em_wo_gas}}
\end{figure}

\begin{figure}[htpb!]
\centering
\includegraphics[width=9.5cm]{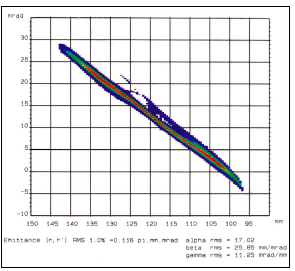}
\caption{Emittance picture after injecting $^{84}$Kr into the beam line:
$\epsilon_{\rm {rms}} = 0.11${\u}$\rmpi${\u}mm{\u}mrad, $p_1=3.5\times10^{-5}${\u}Torr, $p_2=2.7\times10^{-5}${\u}Torr. \label{em_w_Kr}}
\end{figure}

Figures \ref{em_wo_gas}  and \ref{em_w_Kr}  show the emittance measurement obtained
with $^{84}$Kr. It must be pointed out that the emittance
pictures become straight with the gas injection and present a
lower beam size and lower aberrations. This is also evident
during the measurement because each beamlet selected by
the sampler has a lower width and a higher intensity.

\begin{figure}[htpb!]
\begin{center}
\includegraphics[width=0.60\textwidth]{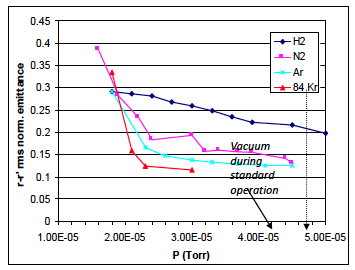}
\end{center}
\caption{Space-charge compensation with H$_2$, N$_2$, Ar and Kr\label{all_gas}}
\end{figure}

\begin{figure}[htpb!]
\begin{center}
\includegraphics[width=0.60\textwidth]{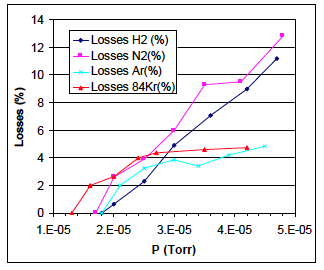}
\end{center}
\caption{Losses of the beam current\label{all_gas_losses}}
\end{figure}

Figure~\ref{all_gas} summarizes the results obtained and in all cases a decrease of beam emittance has been observed
with the increase of beam line pressure.
Figure \ref{all_gas_losses} summarizes also the losses at the end of the beam line at the different pressures, losses that are less than 5\% for masses heavier than Ar.

\subsection{TRIPS beam emittance improvements}

The requirement in terms of proton current for the TRIPS source (Table \ref{accelerator}) was significantly lower than that for the SILHI source, and therefore (at the nominal beam current) gas injection has been not necessary to achieve the required goal. However, also in this case this method has been important to decrease the emittance at higher currents.

In this case the measured values were coherent with the
calculated ones and close to those predicted by the equation
\begin{equation}
\label{freqcollgen}
\epsilon_{\rm{normal,rms}}=\frac{1}{4}\times0.0653r\left(\frac{kT}{A}\right)^{1/2}  \qquad\quad (\rmpi{\u}\mathrm{mm}{\u}\mathrm{mrad}),
\end{equation}
where $r$ is the radius of the extraction aperture in millimetres, $A$ is the rest
mass of the ion in atomic units and $kT$ is the ion temperature in
electronvolts. Assuming an ion temperature of 1{\u}eV and with an extraction hole of 3{\u}mm radius, the expected r.m.s.\ normalized emittance is circa 0.05{\u}$\rmpi${\u}mm{\u}mrad. The emittance values measured are close to the theoretical one for a beam current close to the nominal value, while at higher currents the space-charge effects play an important role in the beam blowup, and a gas must be added into the beam line to decrease the beam emittance.

Measurements \cite{cel2} have been carried out at different current
levels (between 30 and 50{\u}mA); for a
fixed current value, we have explored the effect of some parameters
such as puller voltage and solenoid current on the beam emittance.
A first observation was that the emittance slightly
changes with the puller voltage for a fixed current. Measurements
performed on a 42{\u}mA beam, with the solenoid fixed
at 290{\u}A, a hydrogen flux of 0.49{\u}sccm and a discharge
power of 650{\u}W, show a decrease of emittance from 0.208 to
0.172{\u}$\rmpi${\u}mm{\u}mrad just by increasing the voltage drop between
the extraction and puller electrode from 38 to 43{\u}kV.
The optimal operating value ranges between
40 and 42{\u}kV; unfortunately, for higher voltage HV
discharges occur and it was not possible to operate the source safely.
Another important observation was obtained by looking
at the evolution of the beam emittance by changing the solenoid
strength for fixed source conditions.

\begin{figure}[htpb]
\begin{center}
\includegraphics[width=0.50\textwidth]{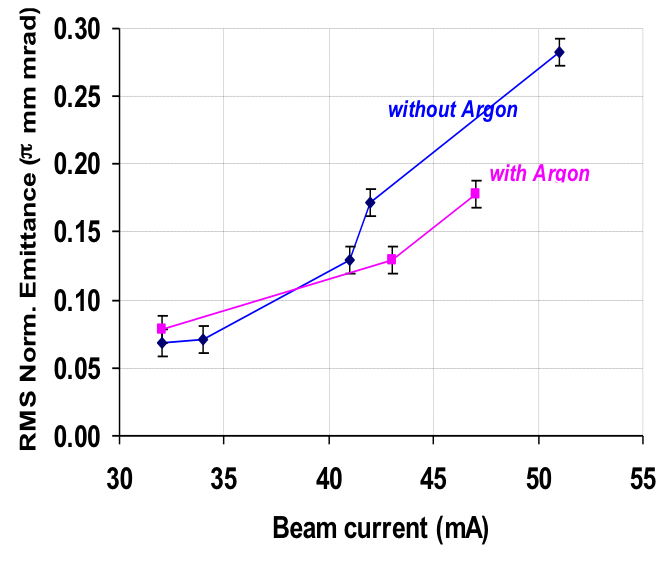}
\end{center}
\caption{The r.m.s.\ normalized emittance versus beam current\label{emit_meas}}
\end{figure}

Measurements were carried out on a 32{\u}mA beam with the puller
fixed at 40{\u}kV, the discharge power at 550{\u}W and the hydrogen
flux at 0.45{\u}sccm; the emittance changes between 0.069
and\break 0.137{\u}$\rmpi${\u}mm{\u}mrad by decreasing the solenoid current.
This in turn means that the position of the crossover along
the beam line is extremely important for the optimum coupling
with the RFQ, as at the crossover the space-charge
forces are more important, as confirmed by the measurements with a four-grid analyser.
An optimal value of solenoid \ced{current} for the operation with the
EMU around 280{\u}A has been found to avoid an excessive
power density over the EMU sampler. Finally Fig.~\ref{emit_meas} shows the evolution of the emittance with the beam current increase; \ced{this confirms} the
influence of beam line pressure on the space-charge compensation (and thus on the emittance) by means of the injection
of a controlled amount of argon into the beam line, as already
done for the SILHI source for different gases.

\begin{figure}[htpb]
\begin{center}
\includegraphics[width=0.45\textwidth]{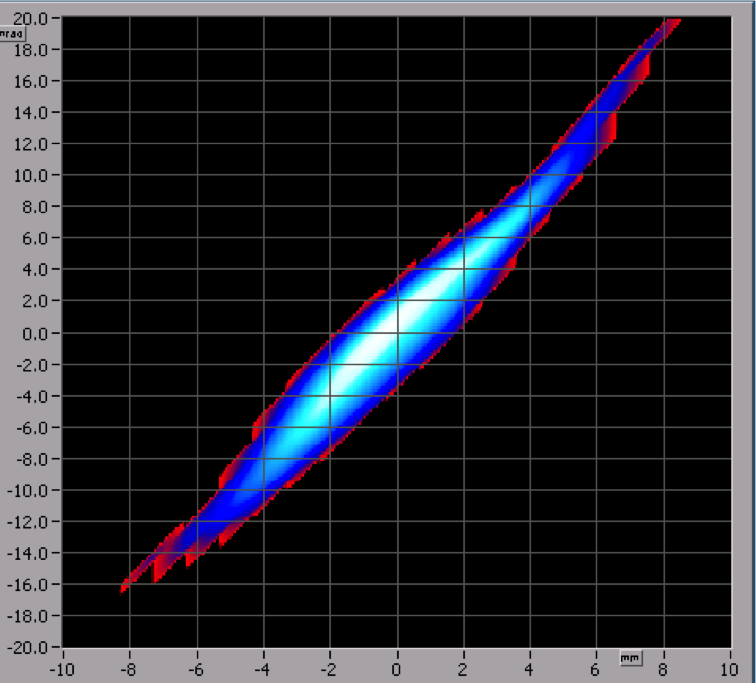}
\end{center}
\caption{Proton beam emittance measured without gas ($P_{\rm source}
= 1.5 \times 10^{-5}${\u}mbar, $P_{\rm line} = 8.8 \times 10^{-6}${\u}mbar,
$I_{\rm extr} = 32{\u}${\u}mA, $\epsilon_{\rm norm} = 0.069${\u}$\rmpi${\u}mm{\u}mrad). \label{color_emit}}
\end{figure}

The measures have been carried out for fixed solenoid
current (280{\u}A) and puller voltage (40{\u}kV) and by changing
the discharge power from 450 to 650{\u}W and the hydrogen
flux from 0.45 to 0.59{\u}sccm. It has been observed that the
space-charge forces play a role for currents greater than
40{\u}mA; the background pressure in that condition was
around $1.5 \times 10^{-5}${\u}mbar in the extraction column and around
$8.5 \times 10^{-6}${\u}mbar in the beam line. By adding an argon leak, the
pressure increased to $2.1 \times 10^{-5}${\u}mbar in the whole line, and
a 30\% beam emittance decrease was obtained without
significant beam losses. 

\begin{figure}[htpb]
\begin{center}
\includegraphics[width=0.45\textwidth]{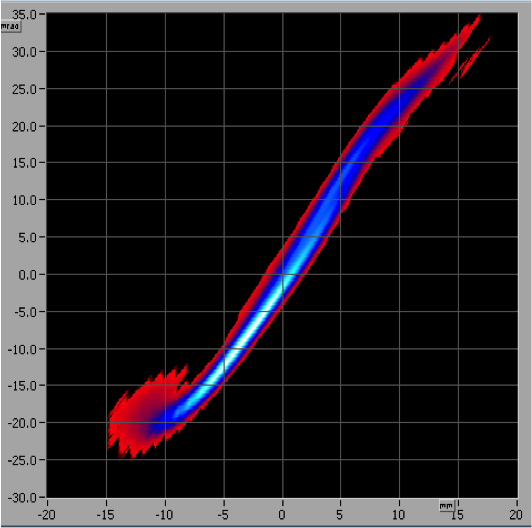}
\end{center}
\caption{Proton beam emittance measured without Ar ($P_{\rm source} =
1.9\times10^{-5}${\u}mbar, $P_{\rm line}=1.1\times10^{-6}${\u}mbar,
$I_{\rm extr}=51${\u}mA, $\epsilon_{\rm norm}=0.283${\u}$\rmpi${\u}mm{\u}mrad).
\label{emit_patt}}
\end{figure}

\begin{figure}[htpb]
\begin{center}
\includegraphics[width=0.45\textwidth]{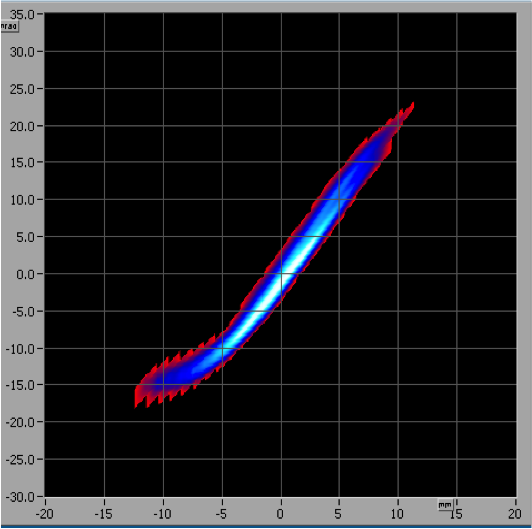}
\end{center}
\caption{Proton beam emittance measured with Ar ($P_{\rm source} =
2.45\times10^{-5}${\u}mbar, $P_{\rm line}=2.2\times10^{-5}${\u}mbar,
$I_{\rm extr}=47${\u}mA, $\epsilon_{norm}=0.178${\u}$\rmpi${\u}mm{\u}mrad). \label{em_patt2}}
\end{figure}

Figure \ref{color_emit} shows the emittance measured
for the nominal current, while Figs.~\ref{emit_patt} and \ref{em_patt2} show the emittance patterns for higher currents with and without gas injection.

In these cases the emittance growth is justified because of the lower space-charge compensation; then the benefits of argon injection are
visible. In fact such gas injection improves the compensation and leads to lower value of emittance.

\section{Future perspectives}

\begin{figure}[htpb]
\begin{center}
\includegraphics[width=0.60\textwidth]{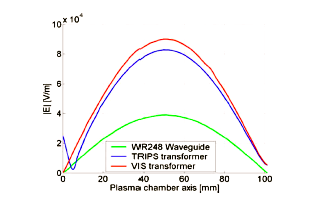}
\end{center}
\caption{Comparison of the electric field on the plasma chamber
axis in different cases\label{comparison}}
\end{figure}

In the off-resonance microwave ion sources, the ECR is not a predominant condition for
plasma generation; in fact, in this case higher electron densities
can be obtained by means of a microwave discharge
at higher magnetic field value and higher pressures. The
plasma--cavity system is a distributed parameter resonant circuit,
since the length of the plasma is of the same order of
magnitude as the free-space electromagnetic wavelength.
Then, differently from the ECR ion sources, the cavity diameter
and excitation frequency are chosen to allow only one
cavity mode to be excited for a given cavity length. In this
case the power coupling into a given mode is usually accomplished
by means of tuning stubs in order to achieve the
necessary impedance match. The steady-state microwave
discharge is characterized by the equality between the power
absorbed by the plasma and the lost power, mainly due to
inelastic ionization, excitation collisions and energy transmission
out of the active discharge region. The power absorbed
by the plasma is given by one-half of the real part of the
complex Poynting vector, and therefore it depends on the electric
field strength in the plasma. In order to optimize the
coupling for a given mode, \ced{the maximization
of that electric field is therefore important}. A waveguide transformer (usually
maximally flat design) is widely used for such purposes
for most microwave sources in operation nowadays.
Such a device realizes a progressive matching between the
waveguide, normally operating in the dominant mode, and
the equivalent impedance of the plasma-filled chamber, also
concentrating the electric field at its centre. Figure \ref{comparison} shows a
comparison of the electric field on the plasma chamber axis in
the case that no transformer is used or by employing the
TRASCO Intense Proton Source (TRIPS) or Versatile Ion Source
(VIS) \cite{celgam}. The excitation frequency is 2.45{\u}GHz in all
cases and it can be observed that an increase by a factor of~2
can be obtained, in the frequency range of circa 400{\u}MHz, by
appropriately shaping the waveguide ridges as discussed in detail
elsewhere~\cite{mai2}.

Therefore, both matching transformers
concentrate the electric field around its axis in a smaller region
than the original WR~284 cross-section. This feature has
been observed in the traces on the boron nitride disc at the
injection side of the TRIPS and VIS plasma chambers, and it is
particularly remarkable for the production of ion beams because
the extraction system is centred on the axis of the
plasma chamber and any enhancement of the plasma density
at the centre of the cavity \ced{leads} to a similar increase in the
ion beam current. Even higher enhancements are made possible
with such a device with different designs or by extending
the ridge also within the span of the plasma chamber. Nowadays
most of the \ced{high-intensity ion sources use} automatic
tuning units to optimize the power coupling into the operational
mode and waveguide transformers similar to those
previously described to enhance the plasma density inside
the source. The increase in the latter parameter is mandatory
for further increase in the produced currents.

Usually in microwave discharge ion sources used for intense
beam production, the microwaves are provided by means of
a waveguide located longitudinally with respect to the chamber
axis. In these devices, the axis of symmetry of the magnetic
field coincides with the chamber axis; this means that
the injected wave is mainly an O or R wave when it propagates
inside the plasma. However, because of the complex
structure of the magnetic field lines, and because of possible
reflections at the chamber walls, these modes may convert
each other, and also X modes can be generated somewhere
inside the cavity, through an O--X conversion. At that time,
the X waves can be directly converted into a Bernstein
electrostatic wave (BW) at the \ced{upper hybrid resonance}
(UHR) layer.  \ced{BWs are a great advantage, as they
travel into the plasma without any cutoff and
provide heating even in the case} of overdense plasmas. However,
if the injection angle of the O mode is not \ced{optimal},
the BW creation process has a low efficiency. The variations
of the power and the background pressure change the
plasma properties, and in turn affect the conversion efficiency.
As BWs are known to be absorbed at cyclotron harmonics,
the electrostatic wave generation
in an MDIS is possible and can be used as an alternative plasma
heating method. In order to increase the BW creation efficiency,
the proper injection angle is needed: this can be
achieved by using single cut antennas (waveguides) launching
O waves in \ced{the right} direction with respect to the magnetic
field lines. In this way O--X--B conversion is possible,
as observed in Ref.~\cite{laqua}. More details are discussed in
Ref.~\cite{laqua2}. A detailed investigation of this new
approach to plasma heating in an MDIS may \ced{make it possible
to take} a large step towards higher extracted beam
currents: in fact the key parameter in MDISs is the electron
density, more \ced{so} than the temperature (low charged ions are
usually required), and by means of electrostatic wave heating
we should be able to overcome the cutoff density at 2.45{\u}GHz
of a factor of 5 or 6.

\newpage

\end{document}